\documentclass[11pt,twoside]{article}
\usepackage{asp2010}

\resetcounters

\begin{document}

\title{Rubidium-rich Asymptotic Giant Branch stars in the Magellanic Clouds}
\author{D. A. Garc\'\i a-Hern\'andez$^{1,2}$
\affil{$^1$Instituto de Astrof\'{\i}sica de Canarias, C/ Via L\'actea
s/n, E$-$38200 La Laguna, Tenerife, Spain; agarcia@iac.es}
\affil{$^2$Departamento de Astrof\'{\i}sica, Universidad de La Laguna (ULL),
E$-$38205, La Laguna, Tenerife, Spain}}

\begin{abstract}
The Magellanic Clouds (MCs) offer a unique opportunity to study the stellar
evolution and nucleosynthesis of massive Asymptotic Giant Branch (AGB) stars in
low metallicity environments where distances are known. Rubidium is a key
element to distinguish between high mass AGB stars and low mass AGBs or other
type of astronomical objects such as massive red supergiant stars.
Theoretically, high mass AGBs are predicted to produce a lot of Rb. We present
the discovery of massive Rb-rich AGB stars in the MCs, confirming for the first
time that these stars also exist in other galaxies. Our findings show that these
stars are generally brighter than the standard adopted luminosity limit
(M$_{bol}$$\sim$$-$7.1) for AGB stars. The observations of massive MC AGBs are
qualitatively predicted by the present theoretical models. However, these
theoretical models are far from matching the extremely high Rb overabundances
observed. This might be related with an incomplete present understanding of the
atmospheres of these stars.
\end{abstract}

\section{Introduction}

The Magellanic Clouds (hereafter MCs) provide a unique opportunity to study the
evolution and nucleosynthesis of low- and intermediate-mass stars (0.8 $<$ M $<$
8 M$_{\odot}$) in low metallicity environments where distances - and hence
luminosity - are known. Low- and intermediate-mass stars experience thermal
pulses and strong mass loss on the Asymptotic Giant Branch \citep[AGB;
e.g.,][]{herw05}. Repeated thermal pulses and ``3$^{rd}$ dredge-up" events can
convert the originally O-rich AGB star into a C-rich one. However, in the case
of the more massive AGB stars (M$>$4 M$_{\odot}$), Hot Bottom Burning
\citep[HBB; e.g.,][]{mazz99} prevents the C-star formation and these stars
remain O-rich despite the dredge-up. The activation of HBB in massive AGB stars
is supported by previous studies on visually bright MC AGB stars
\citep[e.g.,][]{plez93} and on heavily obscured O-rich AGBs - the so-called
``OH/IR" stars - of our Galaxy \citep{gh06,gh07}. 

AGB stars also produce heavy neutron-rich elements ($s$-process elements) such
as Rb, Zr, Sr, Nd, Ba, etc., which can be dredged-up to the stellar surface
\citep[e.g.,][]{buss99}. In the more massive AGB stars, free neutrons are
predicted to be mainly released by the $^{22}$Ne($\alpha$,n)$^{25}$Mg reaction,
while the $^{13}$C($\alpha$,n)$^{16}$O reaction seems to be the dominant neutron
source in lower mass AGB stars. Rb is a key element to distinguish between the
operation of the $^{13}$C versus the $^{22}$Ne neutron source in AGB stars and,
as such, is a good indicator of the progenitor stellar mass\footnote{Note that
other astronomical sources such as massive red supergiant stars are not expected
to overproduce Rb.}. This is because the relative abundance of Rb to other
$s$-process elements such as Zr (i.e., the Rb/Zr ratio) is sensitive to the
neutron density owing to branchings in the s-process path at $^{85}$Kr and
$^{86}$Rb \citep[e.g.,][]{raai08b}. Interestingly, we discovered  strong Rb
overabundances (up to 10$-$100 times solar) with apparently only mild Zr
enhancements in massive galactic O-rich AGB stars \citep{gh06}. This work
provided the first observational evidence that $^{22}$Ne is the dominant neutron
source in the more massive AGB stars. Surprisingly, Rb was not found to be
overabundant in the few unobscured O-rich massive AGBs previously studied in the
SMC \citep{plez93}. Here, we present the first detections of massive Rb-rich AGB
stars in the MCs. 

\section{Optical spectra and abundances}

High-resolution (R$\sim$60,000) optical UVES spectra were obtained for a
carefully selected sample of heavily obscured O-rich AGBs in the MCs
\citep[see][for more details]{gh09}. These stars are suspected to be the most
massive and extreme AGBs known in the Clouds. Sample spectra around the 7800
\AA\ Rb\,{\sc i} line are shown in Fig. 1. We found four Rb-detected AGB stars
in the LMC and one in the SMC (see Fig. 1). 

\articlefigure[angle=-90,scale=.50]{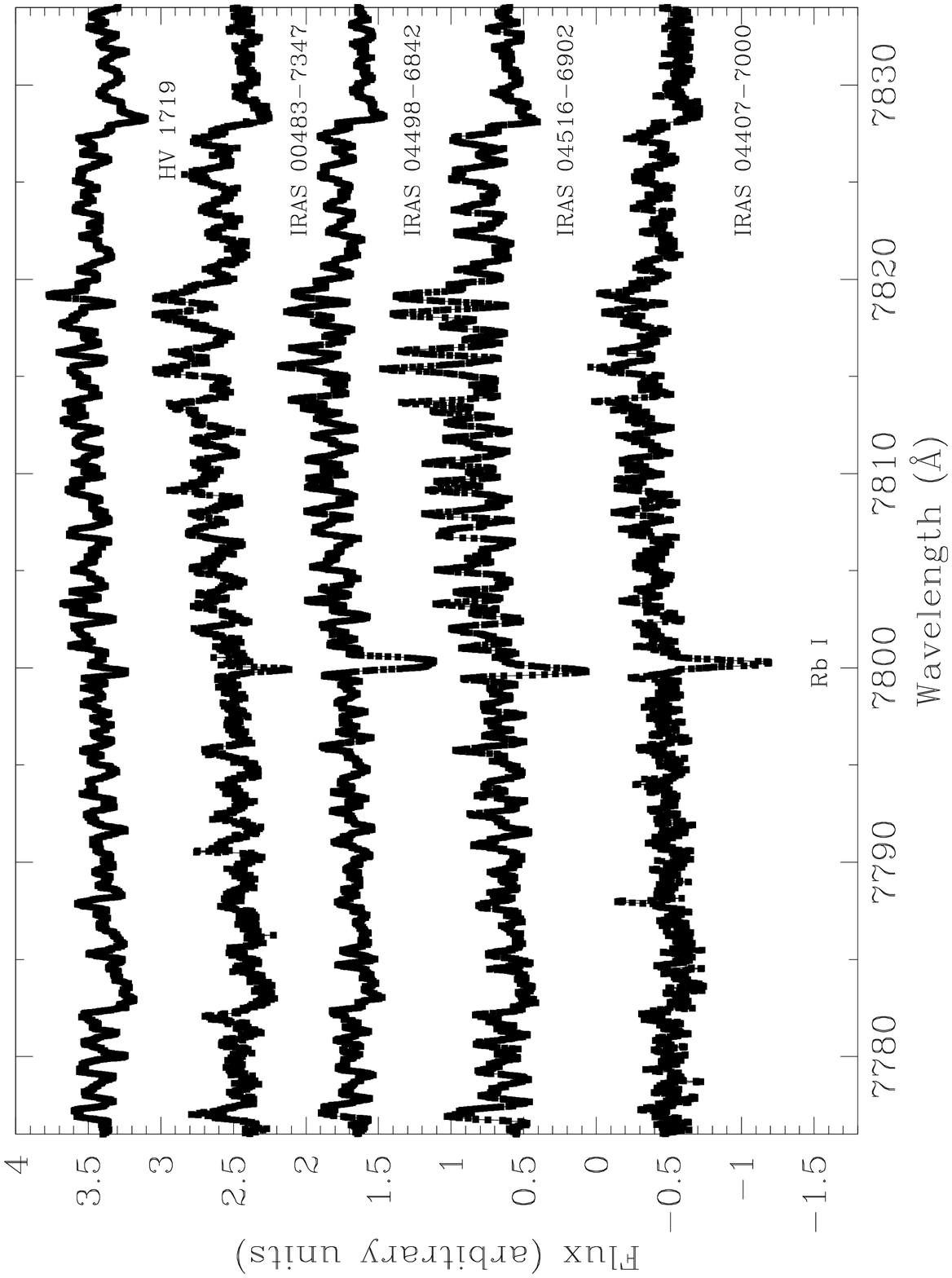}{reference label}{
Sample spectra around the 7800 \AA\ Rb\,{\sc i} line.  The lower three stars
(IRAS 00498-6842, 04516-6902, and 04407-7000) are Rb-rich LMC stars. The upper
two stars are from the SMC. IRAS 00483-7347 is the heavily obscured Rb-rich AGB
discovered in the SMC, while HV 1719 is a visually bright HBB SMC AGB with a
much weaker Rb\,{\sc i} line.}

The abundances (or upper limits) of the elements Rb and Zr (when possible) were
derived from the 7800 \AA\ Rb\,{\sc i} line and the ZrO molecular bands,
respectively, and by following the procedure that we used for the galactic
O-rich AGB stars \citep{gh06,gh07} \citep[see also ][for more details]{gh09}.
The spectroscopic effective temperatures and abundances are summarised in Table
1. The uncertainties of the derived abundances are estimated to be 0.8 and 0.5
dex for Rb and Zr abundances, respectively.  

\begin{table}[!ht]
\caption{Spectroscopic temperatures and abundances for the Rb-rich MC AGBs}
\smallskip
\begin{center}
{\small
\begin{tabular}{ccccc}
\tableline
\noalign{\smallskip}
IRAS name &   T$_{eff}$ & [Rb/z] & [Zr/z]  & Type\\
\noalign{\smallskip}
\tableline
\noalign{\smallskip}
04498$-$6842 & 3400 & $+$5.0		& $\leq$ $+$0.3   & OH/IR \\ 
04407$-$7000 & 3000 & $+$3.2		& $\dots$	  & OH/IR \\
04516$-$6902 & 3000 & $+$3.2$^{*}$	& $\leq$ $+$0.3   & OH/IR? \\ 
05558$-$7000 & 3400 & $+$2.8		& $\dots$	  & OH/IR  \\
00483$-$7347 & 3400 & $+$1.7$^{*}$	& $\dots$	  & OH/IR? \\
\noalign{\smallskip}
\tableline
\end{tabular}
}
\end{center}
\end{table}

\section{Luminous Rb-rich MC AGBs}

The main result of our spectroscopic survey is that we have detected strong
Rb\,{\sc i} lines in AGB stars (four stars in the LMC and one in the SMC) in a
low-metallicity extragalactic system. These first detections of extragalactic
Rb-rich AGB stars confirm that these stars also exist in other galaxies. Similarly
to our Galaxy, the Rb-detected stars in the MCs belong to the type of so-called
OH/IR stars (Table 1). Unfortunately, we could estimate photospheric Rb abundances
in only three LMC Rb-rich stars due to the clear presence of blue-shifted
circumstellar Rb\,{\sc i} lines in the other Rb-detected stars (those stars marked
with an asterisk in Table 1). The extremely high Rb abundances observed (up to
10$^{3}$$-$10$^{5}$ times solar) among the LMC stars are even greater than those
displayed by their Galactic counterparts \citep{gh06}.

A common characteristic of the Rb-rich stars that sets them apart from other AGB
stars is their bolometric luminosity. A plot of M$_{bol}$ vs. [Rb/z] is shown in
Fig. 2. The Rb-rich stars in the LMC are brighter ($-$8$<$M$_{bol}$$<$$-$7) than
the AGB stars with no Rb detected. The jump of the Rb abundances at luminosities
of M$_{bol}$ of $-$7.1 is intriguing (Fig. 2). This bolometric luminosity is the
generally adopted limit for AGB stars \citep{pacz71}. Massive red supergiants
have been thought to be more luminous than the standard limit of
M$_{bol}$$\sim$$-$7.1. Thus, our observations confirm that massive
Rb-rich AGB stars are generally brighter than this limit due to a luminosity
contribution from HBB. HBB models suggest that the Rb-rich LMC AGB stars with
M$_{bol}$ $<$ $-$7 are the descendants of stars with initial masses of at least
$\sim$6$-$7 M$_{\odot}$ \citep[see Fig. 7 in][]{ven00}. 

\articlefigure[angle=-90,scale=.50]{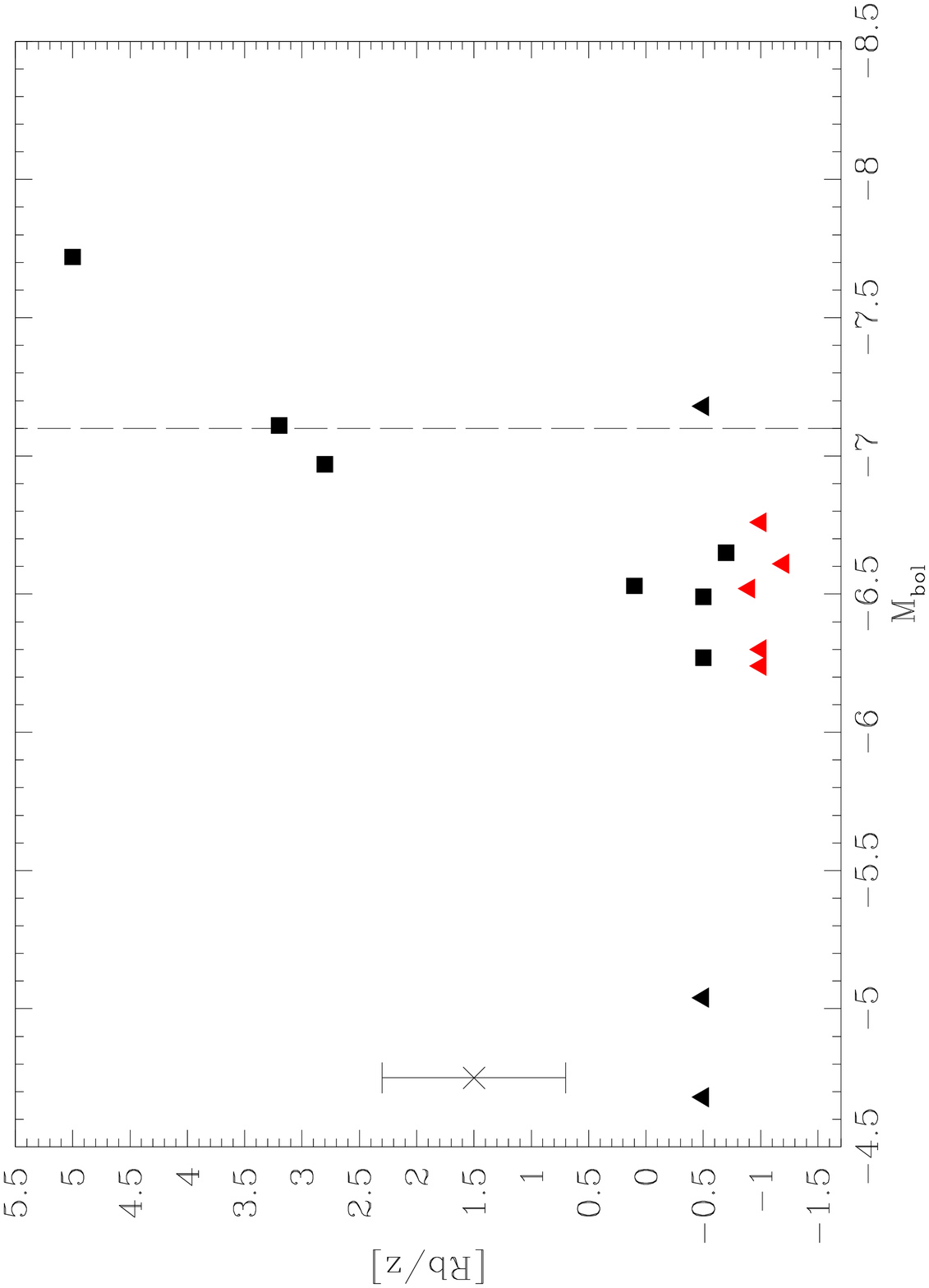}{fig2}{Observed Rb
abundances (squares and triangles for the LMC and SMC, respectively) versus
M$_{bol}$. The Li-rich HBB-AGBs previously studied in the SMC \citep{plez93} are
also included. A typical error bar of $\pm$0.8 dex is shown. The dashed vertical
line marks the theoretical luminosity limit \citep[M$_{bol}$=$-$7.1;][]{pacz71}
for AGB stars.}

\section{Models vs. observations: a Rubidium problem}

We found a Rb problem that is mainly posed by the the four Rb-rich LMC AGB
stars. This Rb problem has two parts: the extremely high Rb abundance and the
extraordinary [Rb/Zr] ratio. Our detections of Rb-rich LMC AGB stars is assured
by inspection of our spectra (Fig. 1). However, the strong Rb overabundance
([Rb/z]$ \sim  +2.8$ to $+5.0$) may be somewhat uncertain because the Rb\,{\sc
i} line is strong and saturated with possible circumstellar contamination. In
addition, our upper limit to the Zr abundance that gives the
extraordinary [Rb/Zr] ratio of $\>$3 to 4 (Table 1) comes from a fit to ZrO
bands.

The Rb overabundance is very likely the consequence of the $s$-process
nucleosynthesis via the high neutron density $^{22}$Ne neutron source, which is
expected to be efficiently activated in massive AGB stars. Although the increase
in the Rb abundance between the low and high neutron density $s$-process paths
(i.e., $^{13}$C vs. $^{22}$Ne) is about an order of magnitude, the predicted
Rb/Zr ratios, however, do not reach extreme values. Present theoretical
predictions for massive AGB models at the LMC and SMC metallicities computed for
the $^{22}$Ne neutron source are far from matching the extremely high Rb
enhancements and the extraordinary [Rb/Zr] ratios (Table 1) that we observe
\citep{gh09}. Thus, within the framework of the s-process it is not possible to
produce extremely high Rb abundances without co-producing Zr at similar levels.
However, these massive AGB nucleosynthesis models can qualitatively describe the
observations of Rb-rich AGBs in the sense that increasing Rb overabundances with
increasing stellar mass and with decreasing metallicity are theoretically
predicted \citep{raai08a,raai08b}. 

The extremely high Rb enhancements and the extraordinary [Rb/Zr] values could
suggest that a nucleosynthesis process not predictable yet is at work in these
massive Rb-rich AGB stars. However, a failure of the adopted models to represent
the real stars may explain the large discrepancy between models and
observations. In addition, the use of the ZrO bands to estimate the Zr abundance
as well as non-LTE effects may be contributing factors too. More realistic models
(e.g., the inclusion of a circumstellar envelope or the application of dynamical
models) would help to solve the discrepancy observed. Despite the uncertainties
in the Rb abundance determinations, the fact that Rb-rich AGB stars  are found
among the most luminous AGB stars (i.e., confined to bolometric magnitudes
M$_{bol} = -7.1$ and brighter; Fig. 2) is assured. This result will help to
identify these stars in other galaxies of the Local Group. 

\acknowledgements D.A.G.H. acknowledges support for this work provided by the
Spanish Ministry of Science and Innovation (MICINN) under the 2008 Juan de la
Cierva Programme and under grants AYA$-$2007$-$64748 and AYA$-$2008$-$04874.
This work is based on observations made at ESO, 080.D-0508(A).

\bibliography{asp_garcia_hernandez}

\end{document}